\documentclass[aps,preprint,showpacs,showkeys]{revtex4}
\usepackage{amsfonts}
\usepackage{subfigure}
\usepackage{amsmath}
\usepackage{amssymb}
\usepackage{graphicx}
\usepackage{epsfig}
\usepackage{lineno}

\input{tcilatex}

\begin{document}

\preprint{}
\title[ ]{Relativistic quantum speed limit time in dephasing noise}
\author{Salman Khan$^{\dag}$}
\email{sksafi@comsats.edu.pk}
\author{ Niaz Ali Khan$^{\ddag }$}
\affiliation{$^{\dag }$Department of Physics, COMSATS Institute of Information
Technology,Park road, Chak Shahzad, Islamabad, Pakistan, 45550}
\affiliation{$^{\ddag}$CFP and Departamento de F\'{\i}sica, Faculdade de Ci\^{e}ncias,
Universidade do Porto, 4169-007 Porto, Portugal.}
\date{August 10, 2015}

\begin{abstract}
The behavior of quantum speed limit time (QSLT) for a single free spin $-1/2$
particle described by Gaussian wavepackets in the framework of relativity
under dephasing noise is investigated. The dephasing noise acts only on the
spin degrees of freedom of the spin$-1/2$ particle. In particular, the
effects of initial time parameter, rapidity, average momentum and the size
of the wavepackets in the presence of the dephasing noise on the dynamics of
evolution process are studied. In general, the effects of relativity
monotonically decrease the QSLT in time. In the range of large values of
average momentum, critical values of both the rapidity and the size of the
wavepackets exist at which the QSLT has its minimum value. In the range of
small values of the average momentum, the QSLT monotonically decreases with
both rapidity and the size of the wavepackets. The decrease of QSLT in a
particular range of rapidity and with other relative parameters may be of
great interest in employing fast quantum communication and quantum
computation.
\end{abstract}

\maketitle

\section{Introduction}

In quantum information theory entanglement plays a vital role due to which\
investigating its dynamics under variety of situations has been the focus
for a long time \cite{Ekert}. The study of Quantum information processing in
the framework of relativity is a challenging task and is presently under
exploration. Initially, focused on the study of the dynamics of bipartite
entanglement under different circumstance \cite%
{Peres,Soo,Alsing,Fuentes,Khan1}, the theoretical analysis of its effects to
other scenarios have also been lately extended \cite{Single,Khan2, Khan3}.
The output of such studies is an established fact that entanglement is a
relative quantity and degrades with the acceleration of the observer frame.
Now the question arises whether its only the entanglement that depends on
the frame of reference or others quantum mechanical properties of a quantum
system also behave this way. Keeping this in mind, we wish to investigate
the effect of relativity on quantum speed limit time (QSLT) of a single spin$%
-1/2$ particle described by Gaussian wavepackets in the presence of
dephasing noise that acts only on its spin degrees of freedom.

The evolution of a quantum system from an initial state to one of its
allowed orthogonal states is not instantaneous rather the laws of quantum
mechanics put a speed limit to the evolution within the Hilbert space of the
system. The role of such limits has already been studied in different
setups, such as, the identification of precision bounds in quantum metrology 
\cite{Giovanetti}, quantum computation \cite{Bekenstein}, the formulation of
computational limits of physical systems \cite{Lloyd} and the development of
quantum optimal control algorithms \cite{Caneva}. The QSLT is the minimum
time a quantum system takes to evolve within its Hilbert space from one to
another allowed state. Different lower bounds on QSLT for isolated system
have been obtained \cite{Mandelstam,Vaidman,Morgolus,Levitin} and are then
extended to nonorthognal states as well as to driven systems \cite%
{Giovanetti2,Jones,Deffner}. The QSLT of a pure initial state as an open
system for a given driving time and with non-Morkovian dynamics has also
been investigated \cite{Taddei,Campo,Deffner2}. Recently, based on the
relative purity as the distance measure, a scheme has been developed to
measure QSLT for the nonunitary evolution of open mixed initial state \cite%
{Zhang}.

In this paper we use the approach of Ref. \cite{Zhang} and find the effects
of different parameters on the QSLT of the spin degrees of freedom of spin $%
-1/2$ particle in the frame work of relativity. We consider that the system
is initially in a coherent superposition state and is coupled to an
Ohmic-like dephasing environment. Our investigation focuses on the dynamics
of QSLT both under Ohmic and super-Ohmic dephasing environment. We show that
the effect of the relative motion of the observer on the evolution process
of the system is alike irrespective of the initial coherence in the state of
the system. Our findings show that in the Markovian regime the relative
motion of the detector speeds up the evolution process in time. There exists
a critical value $\alpha _{c}$ of the rapidity in the limit of large
normalized averaged momentum at which the speed of the evolution process is
maximum both under the action of Ohmic and super-Ohmic reservoirs. Beyond $%
\alpha _{c}$, it slows down until it reaches a saturation value and becomes
static. It is found that at constant rapidity a similar behavior exists
against the width $W$ of the wavepacket.

\section{Theoretical model}

We consider the time evolution of the spin degrees of freedom of a spin $%
-1/2 $ particle (qubit), observed by a moving detector, interacting with a
dephasing environment. Such a scenario can well be described in terms of the
following Hamiltonian \cite{Addis,Fanchini,Crow}%
\begin{equation}
H=\frac{1}{2}\omega _{0}\sigma _{z}+\sum_{j}\omega _{j}a_{j}^{\dag
}a_{j}+\sigma ^{z}\sum_{j}(g_{j}a_{j}^{\dag }+g_{j}^{\ast }a_{j}),
\label{E1}
\end{equation}%
where the first and the second terms describe the independent evolution of
the qubit and the environment, respectively. The third term of the
Hamiltonian describes the interaction between the qubit and the environment
whose strength with the $j$th mode of the field of frequency $\omega _{j}$
is specified by the constant $g_{j}$. The creation and annihilation
operators of the $j$th mode of the environment obey the usual commutation
relation $\left[ a_{j},a_{j^{\prime }}^{\dag }\right] =\delta _{j,j^{\prime
}}$. In Eq. (\ref{E1}), the $\omega _{0}$ and $\sigma _{z}$, respectively,
represent the transition frequency and the evolution operator of the qubit.
It is easy to check that Eq. (\ref{E1}) commutes with $\sigma _{z}$ thereby
limiting the off-diagonal elements of the density operator, describing the
coherence of the system, to zero and hence leaving the population terms
unchanged. We begin from a factorized initial composite state of the qubit
and the environment with the environment being in its vacuum state at zero
temperature. In the limit of very large environment, the sum over the
discrete coupling constants $g_{j}$ between the different modes of the
environment and the qubit can be replaced by an integral over a continuous
distribution of frequencies of the environment modes, that is, $%
\sum_{j}\left\vert g_{j}\right\vert ^{2}\rightarrow \int_{0}^{\infty
}J(\omega )d\omega $, where $J(\omega )$ stands for the spectral density of
the environmental modes. For an Ohmic like dephasing model, it can be
expressed as follows%
\begin{equation}
J(\omega )=\frac{\omega ^{n}}{\omega _{c}^{n-1}}\eta e^{-\omega /\omega
_{c}},  \label{E2}
\end{equation}%
where $\omega _{c}$ is the cutoff frequency and $\eta $ is a dimensionless
coupling constant. The spectral density is called sub-Ohmic, Ohmic and
super-Ohmic for $n<1,$ $n=1$ and $n>1$, respectively. We will limit our
analysis only to the last two cases.

The evolved state of an open qubit at any time $t$ can be written in terms
of Kraus operators as follows%
\begin{equation}
\rho (t)=\sum_{k=1}^{2}E_{k}\rho (\tau )E_{k}^{\dag },  \label{TEvolution}
\end{equation}%
where $\rho (\tau )$ is the initial density matrix of the system at any
initial time $\tau $ and $E_{k}$ are the single qubit Kraus operators which
are given by%
\begin{equation}
E_{1}=|0\rangle \langle 0|+p_{t}|1\rangle \langle 1|,\qquad E_{2}=\sqrt{%
1-p_{t}^{2}}|1\rangle \langle 1|,
\end{equation}%
where $p_{t}=e^{-\gamma (t)}$ with $\gamma (t)$ being the dephasing rate and
is given by%
\begin{equation}
\gamma (t)=\eta \omega _{c}\Gamma (n)\int_{0}^{t}\sin [n\arctan (\omega
_{c}t^{^{\prime }})](1+(\omega _{c}t^{^{\prime }})^{2})^{-n/2}dt^{^{\prime
}},
\end{equation}%
where $\Gamma \left( n\right) $ is the Euler Gamma function . For an Ohmic
reservoir, it takes the form $\gamma (t)=\eta \ln (1+(\omega _{c}t)^{2})$.
Also, the nonunitary generator of the reduced dynamics of the system for the
quantum dephasing channel is given by%
\begin{equation}
\mathcal{\mathit{L}}\rho (t)=\frac{\gamma (t)}{2}[\sigma _{z}\rho (t)\sigma
_{z}-\rho (t)].  \label{E3}
\end{equation}

Next, we specify the state of the qubit as our system. The generic state, in
the momentum representation, of a qubit in the laboratory frame can be
expressed as \cite{Bogolubov,Peres2}%
\begin{equation}
\left\vert \psi (\mathbf{p})\right\rangle =f_{\mathbf{k}}^{w}(\mathbf{p})%
\begin{pmatrix}
\cos \theta \\ 
\sin \theta%
\end{pmatrix}%
,  \label{E4}
\end{equation}%
where $\theta =[0,\pi /4]$ and describes the initial level of coherence
between the two allowed states, $f_{\mathbf{k}}^{w}(\mathbf{p})=\pi
^{-3/4}w^{-3/2}\exp [-(\mathbf{p}-\mathbf{k})^{2}/2w^{2}]$ with $w\in 
\mathbf{%
%TCIMACRO{\U{211d} }%
%BeginExpansion
\mathbb{R}
%EndExpansion
}_{+}$ is a measured of the dispersion in momentum and $\mathbf{k=(k,}0,0%
\mathbf{)}$ represents the average momentum. In the proper frame of the
detector, which is moving with relativistic speed $v$ with respect to the
laboratory frame, the state of the qubit transforms through a unitary
transformation given by \cite{Halpern,Weinberg}%
\begin{equation}
\left\vert \psi (\mathbf{p})\right\rangle \rightarrow \left\vert \varphi (%
\mathbf{p})\right\rangle =U(\Lambda )\left\vert \psi (\mathbf{p}%
)\right\rangle ,  \label{E5}
\end{equation}%
with%
\begin{equation}
U(\Lambda )\left\vert \psi (\mathbf{p})\right\rangle =\sqrt{\frac{q_{0}}{%
p_{0}}}D\left( \Lambda ,\Lambda ^{-1}p\right) \left\vert \psi (\mathbf{p}%
)\right\rangle .  \label{E6}
\end{equation}%
where $D\left( \Lambda ,\Lambda ^{-1}p\right) $ is the well known Wigner
rotation and is explicitly given by%
\begin{equation}
D\left( \Lambda ,\Lambda ^{-1}p\right) =\frac{\left( p_{0}+m\right) \sigma
_{0}\cosh \left( \alpha /2\right) +(\left( \mathbf{q\cdot e)}\sigma
_{0}+i\left( \mathbf{e\times q}\right) \cdot \mathbf{\sigma }\right) \sinh
\left( \alpha /2\right) }{\left[ \left( p_{0}+m\right) \left( q_{0}+m\right) %
\right] ^{1/2}}.  \label{E7}
\end{equation}%
In Eq. (\ref{E7}), $\alpha =-\tanh ^{-1}(v/c)$ stands for the rapidity, $%
\mathbf{e=}\left( e_{x},0,0\right) $ is the direction of the boost, $\mathbf{%
q=}\left( q_{x},q_{y},q_{z}\right) $ gives the spatial part of the
four-vector with $q=\Lambda ^{-1}p$, $\sigma _{0}$ and $\mathbf{\sigma =}%
\left( \sigma _{x},\sigma _{y},\sigma _{z}\right) $ are $2\times 2$ identity
and Pauli matrices, respectively. Note that from this point onward we will
set $c=1$. For the boost in the $x$ direction, we have%
\begin{equation}
\Lambda =%
\begin{bmatrix}
\cosh \alpha & \sinh \alpha & 0 & 0 \\ 
\sinh \alpha & \cosh \alpha & 0 & 0 \\ 
0 & 0 & 1 & 0 \\ 
0 & 0 & 0 & 1%
\end{bmatrix}%
.
\end{equation}%
The explicit form of the transformed wavefunction of the qubit in the
detector frame is obtained through back substitution of Eq. (\ref{E7}) into
Eq. (\ref{E6}), the result of which along with Eq. (\ref{E4}) into Eq. (\ref%
{E5}) and is given by

\begin{equation}
\left\vert \varphi (\mathbf{p})\right\rangle =\cos \theta \binom{a_{1}\left( 
\mathbf{p}\right) }{a_{2}\left( \mathbf{p}\right) }+\sin \theta \binom{%
-a_{2}\left( \mathbf{p}\right) }{a_{1}\left( \mathbf{p}\right) ^{\ast }},
\label{IState}
\end{equation}%
with%
\begin{eqnarray}
a_{1}\left( \mathbf{p}\right) &=&kf_{\mathbf{k}}\left( \mathbf{q}\right) %
\left[ C\left( q_{0}+m\right) +S\left( q_{x}+iq_{y}\right) \right] ,  \notag
\\
a_{2}\left( \mathbf{p}\right) &=&kf_{\mathbf{k}}\left( \mathbf{q}\right)
Sq_{z},  \notag \\
k &=&\left[ \left( \frac{q_{0}}{p_{0}}\right) /\left( \left( q_{0}+m\right)
\left( p_{0}+m\right) \right) \right] ^{1/2},
\end{eqnarray}%
where the asterisk shows complex conjugate and the new parameters are
defined as $C=\cosh \alpha /2$, and $S=\sinh \alpha /2$.

In order to investigate the behavior of QSLT for our system, we first want
to review the main results for the different measures of the bounds of QSLT
both for pure and mixed initial states proved in \cite%
{Mandelstam,Vaidman,Morgolus,Levitin,Zhang}. Especially, we focus our review
on the recent measure of QSLT \cite{Zhang} derived in terms of relative
purity $f(t)$ between the initial and the final states of a quantum system
which is given as%
\begin{equation}
f(t)=\frac{tr\left( \rho (t)\rho (\tau )\right) }{tr\left( \rho (\tau
)^{2}\right) },  \label{RF}
\end{equation}%
where $\rho (\tau )$ is the initial density matrix at a time $\tau $, and $%
\rho (t)$ represents the density matrix at a later time $t=\tau +\Delta \tau 
$ with $\Delta \tau $ being the driving time. Differentiating Eq. (\ref{RF})
with respect to $t$ and using $\dot{\rho}(t)=\mathcal{L}\rho (t)$ with $%
\mathcal{L}$ being the superoperator representing the nonunitary reduced
dynamics of the system, we get%
\begin{equation}
\dot{f}(t)=\frac{tr(\mathcal{L}\rho (t)\rho (\tau ))}{tr(\rho (\tau )^{2})}.
\label{RF2}
\end{equation}%
If $\lambda _{j}$ stand for the singular values of the initial density
matrix $\rho (\tau )$ and $\mu _{j}$ represent the singular values of $%
\mathcal{L}\rho (t)$, then with the help of von Neumann trace inequality,
the absolute of the numerator of Eq. (\ref{RF2}) can be expressed as follows%
\begin{equation}
\left\vert tr(\mathcal{L}\rho (t)\rho (\tau ))\right\vert \leq
\sum_{j=1}^{m}\lambda _{j}\mu _{j}.  \label{RF3}
\end{equation}%
Since $0<\lambda _{j}\leq 1,$ therefore, $\left\Vert \mathcal{L}\rho
(t)\right\Vert _{tr}=\sum_{j=1}^{m}\mu _{j}\geq $ $\sum_{j=1}^{m}\lambda
_{j}\mu _{j}$, where $\left\Vert \cdot \right\Vert _{tr}$ represents the
trace norm of an operator. With the use of this condition, Eq. (\ref{RF2})
and Eq. (\ref{RF3}) give%
\begin{equation}
\left\vert \dot{f}(t)\right\vert \leq \frac{\sum_{j=1}^{m}\mu _{j}}{tr\left(
\rho (\tau )^{2}\right) }.  \label{RF4}
\end{equation}%
Integrating Eq. (\ref{RF4}) with respect to $t$ between $t=\tau $ and $%
t=\tau +\Delta \tau $, the result can be expressed in the form of the
following inequality%
\begin{equation}
\tau \geq \max \left\{ \frac{1}{\overline{\sum_{j=1}^{m}\lambda _{j}\mu _{j}}%
},\frac{1}{\overline{\sum_{j=1}^{m}\mu _{j}}}\right\} \left\vert f\left(
t\right) -1\right\vert tr\left( \rho (\tau )^{2}\right) .  \label{RF5}
\end{equation}%
In Eq. (\ref{RF5}), the bar over the terms in the denominator represents the
average over time and can be written as%
\begin{equation}
\overline{X}=\frac{1}{\Delta \tau }\int_{\tau }^{\tau +\Delta \tau }X~dt.
\end{equation}%
Using the fact that under unitary transformation, $\overline{%
\sum_{j=1}^{m}\mu _{j}}$ stands for the energy of the system averaged over
time, the Margolus-Levitin (ML) \cite{Morgolus,Levitin} type bound on the
speed of evolution for closed systems from Eq. (\ref{RF5}) can be expressed
as%
\begin{equation}
\tau \geq \frac{\left\vert f\left( t\right) -1\right\vert tr\left( \rho
(\tau )^{2}\right) }{2\overline{\sum_{j=1}^{m}\mu _{j}}}.  \label{RF6}
\end{equation}%
On the other hand, using the inequality $\sum_{j=1}^{m}\lambda _{j}\mu
_{j}\leq \sum_{j=1}^{m}\mu _{j}$ or $\frac{1}{\overline{\sum_{j=1}^{m}%
\lambda _{j}\mu _{j}}}\geq \frac{1}{\overline{\sum_{j=1}^{m}\mu _{j}}}$, Eq.
(\ref{RF5}) modifies to the following form%
\begin{equation}
\tau \geq \frac{\left\vert f\left( t\right) -1\right\vert tr\left( \rho
(\tau )^{2}\right) }{\overline{\sum_{j=1}^{m}\lambda _{j}\mu _{j}}}.
\label{ML}
\end{equation}%
Eq. (\ref{ML}) gives the ML type bound on the speed of nonunitary\ dynamics
of open systems with mixed initial states.

Similarly, taking absolute of Eq. (\ref{RF2}) and employing the
Cauchy-Schwarz inequality for operators, it is straightforward to arrive at
the following result%
\begin{equation}
\left\vert \dot{f}(t)\right\vert \leq \frac{\sqrt{tr\left( \mathcal{L}\rho
(t)^{\dag }\mathcal{L}\rho (t)\right) tr\left( \rho (\tau )^{2}\right) }}{%
tr\left( \rho (\tau )^{2}\right) }.  \label{RF8}
\end{equation}%
For mixed initial state $\rho \left( \tau \right) $, $tr\left( \rho \left(
\tau \right) ^{2}\right) <1$, which implies that $\sqrt{tr\left( \mathcal{L}%
\rho (t)^{\dag }\mathcal{L}\rho (t)\right) tr\left( \rho (\tau )^{2}\right) }%
<\sqrt{tr\left( \mathcal{L}\rho (t)^{\dag }\mathcal{L}\rho (t)\right) }$.
Using this condition, the above equation can be reduced to the following form%
\begin{equation}
\left\vert \dot{f}(t)\right\vert \leq \sqrt{\sum_{j=1}^{m}\mu _{j}^{2}},
\label{RF9}
\end{equation}%
where the use of Hilbert-Schmidt norm given by $\sqrt{tr\left( \mathcal{L}%
\rho (t)^{\dag }\mathcal{L}\rho (t)\right) }=\left\Vert \mathcal{L}\rho
(t)\right\Vert _{hs}=\sqrt{\sum_{j=1}^{m}\mu _{j}^{2}}$ is made. Integration
of Eq. (\ref{RF9}) leads to the following Mandelstam-Tamm (MT) \cite%
{Mandelstam} type bound on QSLT for nonunitary dynamics of quantum systems%
\begin{equation}
\tau \geq \frac{\left\vert f\left( t\right) -1\right\vert tr\left( \rho
(\tau )^{2}\right) }{\overline{\sqrt{\sum_{j=1}^{m}\mu _{j}^{2}}}},
\label{MT}
\end{equation}%
with%
\begin{equation}
\overline{\sqrt{\sum_{j=1}^{m}\mu _{j}^{2}}}=\frac{1}{\Delta t}\int_{\tau
}^{t}\sqrt{\sum_{j=1}^{m}\mu _{j}^{2}}dt,
\end{equation}%
as the time averaged variance of energy. A unified relation for QSLT of
arbitrary mixed initial state interacting with environment can be obtained
by combining Eqs. (\ref{ML}) and (\ref{MT}) as follows \cite{Mandelstam}%
\begin{equation}
\mathcal{T}_{\mathcal{QSL}}=\max \left[ \frac{1}{\overline{\sqrt{%
\sum_{j=1}^{m}\mu _{j}^{2}}}},\frac{1}{\overline{\sum_{j=1}^{m}\lambda
_{j}\mu _{j}}}\right] \left\vert f(t)-1\right\vert tr(\rho (\tau )^{2}).
\label{QSL}
\end{equation}%
Since the singular values $\lambda _{j}$ of a pure initial state $\rho (\tau
)$ obey the condition $\lambda _{j}=\delta _{j,1}$, therefore, $%
\sum_{j=1}^{m}\lambda _{j}\mu _{j}=\mu _{1}\leq \sqrt{\sum_{j=1}^{m}\mu
_{j}^{2}}$. This result is in agreement with the one obtained in \cite%
{Deffner2}.

With all the required tools in hand, we can now use them to find the
dynamics of QSLT for a qubit in the relativistic frame work. The initial
density matrix we start from corresponds to the state given in Eq. (\ref%
{IState}) whose explicitly form becomes%
\begin{equation}
\rho (0)=\frac{1}{2}%
\begin{pmatrix}
1+(1-2\chi )\cos 2\theta & (1-4\chi )\sin 2\theta \\ 
(1-4\chi )\sin 2\theta & 1-(1-2\chi )\cos 2\theta%
\end{pmatrix}%
,
\end{equation}%
where the parameter $\chi =\chi \left( \alpha \right) $ represents the
relativistic effect and is given by \cite{Landulfo}%
\begin{equation}
\chi \left( \alpha \right) =\sinh ^{2}\left( \alpha /2\right) \int \frac{%
q_{z}^{2}}{\left( q_{0}+m\right) \left( p_{0}+m\right) }\left\vert f_{%
\mathbf{k}}^{w}(\mathbf{q})\right\vert ^{2}d\mathbf{q}.  \label{Chi}
\end{equation}%
The analytical solution of Eq. (\ref{Chi}) is difficult, however, we can
solve it numerically by first transforming it into cylindrical coordinates
with $q_{x}$ as the symmetry axis and defining $Q_{r}=q_{r}/m,$ $%
Q_{x}=q_{x}/m,$ $W=w/m,$ $K=k/m$ and $Q_{0}=\sqrt{Q_{r}^{2}+Q_{x}^{2}+1}$ as
the normalized nondimensional variables$.$

The time evolution of the density matrix is obtained by using Eq. (\ref%
{TEvolution}) and can be expressed as%
\begin{equation}
\rho (t)=\frac{1}{2}%
\begin{pmatrix}
1+(1-2\chi )\cos 2\theta & p_{t}(1-4\chi )\sin 2\theta \\ 
p_{t}(1-4\chi )\sin 2\theta & 1-(1-2\chi )\cos 2\theta%
\end{pmatrix}%
.  \label{Fstate}
\end{equation}%
Similarly the nonunitary dynamics of the system are obtained by using Eq. (%
\ref{E3}) and can be written as%
\begin{equation}
\mathcal{\mathit{L}}\rho (t)=\frac{1}{2}%
\begin{pmatrix}
0 & p_{t}(4\chi -1)\gamma (t)\sin 2\theta \\ 
p_{t}(4\chi -1)\gamma (t)\sin 2\theta & 0%
\end{pmatrix}%
.  \label{NonUnitary}
\end{equation}%
In order to investigate the QSLT, we need to find the singular values of
Eqs. (\ref{Fstate}) and (\ref{NonUnitary}). For Eq. (\ref{Fstate}) these are
given as follows%
\begin{equation}
\lambda _{\pm }=\frac{1}{2}\pm \frac{1}{2\sqrt{2}}\sqrt{p_{t}^{2}(1-4\chi
)^{2}+(1-2\chi )^{2}-(p_{t}^{2}(1-4\chi )^{2}-(1-2\chi )^{2})\cos 4\theta }.
\end{equation}%
Similarly, the singular values of Eq. (\ref{NonUnitary}) can be expressed as
follows%
\begin{equation}
\mu _{1}=\mu _{2}=\frac{1}{2}\left\vert \gamma (t)p_{t}(1-4\chi )\right\vert
\sin 2\theta .
\end{equation}%
With the set of eigenvalues given above, the ML type bound is satisfied and
the QSLT for a qubit in the relativistic frame work becomes%
\begin{equation}
\mathcal{T}_{\mathcal{QSL}}=\frac{\left\vert (1-4\chi )\left( p_{\tau
}p_{t}-p_{\tau }^{2}\right) \right\vert }{\Delta \tau ^{-1}\int_{\tau
}^{\tau +\tau _{D}}\left\vert \overset{\centerdot }{p_{t}}\right\vert dt}%
\sin 2\theta .  \label{FinalQSLT}
\end{equation}%
Setting $\chi =0$ and the initial coherence term $\sin 2\theta =[C\left(
\rho _{0}\right) ]^{1/2}$, Eq. (\ref{FinalQSLT}) straightaway goes to the
result of \cite{Zhang}. The presence of $\chi $ in Eq. (\ref{FinalQSLT}),
reveals that relativity affect the QSLT for quantum systems. Since the term
describing the initial coherence factors out from the terms inside the
brackets containing $\chi $, therefore, the effect of relativity on all
initial states, regardless of the degree of initial coherence, is same. One
can easily observe that with the increasing value of $\chi $ and freezing
all the other variables to some constant values, first QSLT decreases
reaching a minimum equal to zero at the critical value of $\chi =1/4$ and
then start increasing. This means that at the critical value of $\chi $ the
evolution of the quantum system is instantaneous. Nevertheless, we believe
that observing this effect is not possible as $\chi $ is itself a function
of other parameters such as $\alpha $, $K$ and $q$ that limit the
instantaneous evolution of quantum system. The different choices of these
parameters give rise to some novel results for QSLT, which we, next,
demonstrate them graphically. In the Markovian regime \cite{Bylicka}, Eq. (%
\ref{FinalQSLT}) reduces to the following form%
\begin{equation}
\mathcal{T}_{\mathcal{QSL}}=p_{\tau }\Delta \tau \left\vert 1-4\chi
\right\vert \sin 2\theta .
\end{equation}%
We use this equation to explicitly further investigate the effects of other
parameters on the dynamics of QSLT by limiting our analysis to the Markovian
regime.

\section{Discussion}

\begin{figure}[h]
\begin{center}
\includegraphics[scale=0.9]{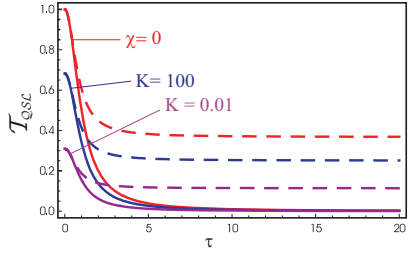}
\end{center}
\caption{(Color Online) The $\mathcal{T}_{\mathcal{QSL}}^{\mathcal{R}}$ as a
function of the initial time parameter $\protect\tau $ by choosing the other
parameters such that $\Delta \protect\tau =1,$ $\protect\eta =1,$ $\protect%
\omega _{c}=1,$ $W=4\ $and $\protect\theta =\protect\pi /4.$ The solid
(dashed) curves correspond to Ohmic (super-Ohmic) reservoir. The red curves
present zero relativistic effect $(\protect\alpha =0)$ whereas the blue and
the purple curves correspond to $K=100$ and $K=0.01$, respectively, in the
limit of infinite rapidity.}
\label{Figure1}
\end{figure}

To get a deep insight into the influence of relativity on the dynamics of
QSLT, we plot it against different parameters under various conditions. In
figure (\ref{Figure1}), we plot the QSLT against $\tau $ in the presence of
environment both for relativistic (infinite rapidity) and nonrelativistic
cases. The solid curves represent its behavior under the influence of Ohmic
reservoir and the dashed curves show its dynamics under the effect of
Super-Ohmic reservoir. One can see that the qualitative behavior of QSLT
under the effect of relativity in both Ohmic and Super-Ohmic limits remains
unchanged. However, quantitatively it is damped such that the damping itself
becomes a function of $K$. For large $K$ the damping is moderate and for
small $K$ it is strong. The notable difference between the effect of the two
limits of the reservoir is that for Ohmic case it goes to zero at different
times depending on the choice of the value of $K$ whereas for super-Ohmic
case it reaches a stable and static value, different for every choice of $K$%
. 

\begin{figure}[h]
\begin{center}
\subfigure[]{
\includegraphics[scale=0.9]{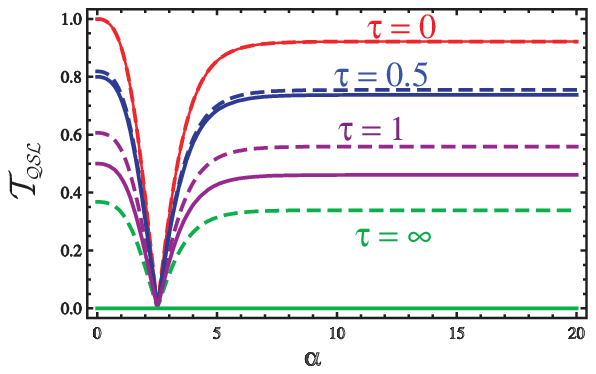}} 
\subfigure[]{
\includegraphics[scale=0.9]{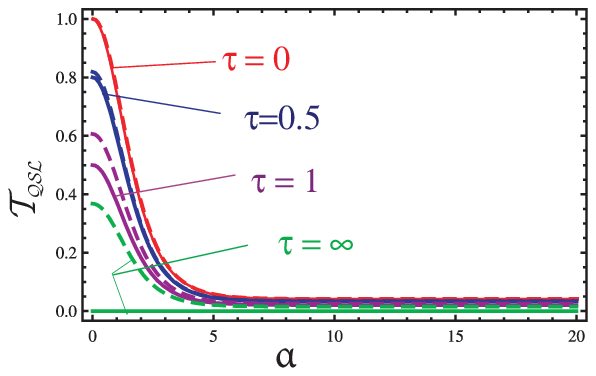}}
\end{center}
\caption{(Color Online) The $\mathcal{T}_{\mathcal{QSL}}^{\mathcal{R}}$ as a
function of rapidity by choosing the other parameters such that $\Delta 
\protect\tau =1,$ $\protect\eta =1,$ $\protect\omega _{c}=1,$ $W=30,$\ $%
\protect\alpha =\infty ,$ $\protect\theta =\protect\pi /4$ and $(a)$ $K=100$%
, $(b)$ $K=0.01.$ The red, the blue, the purple and the green solid (dashed)
curves correspond to the initial time parameter $\protect\tau =0,$ $0.5$ $%
,1, $ $\infty $ for Ohmic(super-Ohmic) reservoir, respectively.}
\label{Figure2}
\end{figure}

The behavior of QSLT against rapidity for different choices of $\tau $ is
shown in figure (\ref{Figure2}$a$). Again the solid curves in both
subfigures represent the effect of Ohmic reservoir and the dashed ones show
the effect of super-Ohmic reservoir. Besides the relatively large damping
caused by the Ohmic reservoir, the effect of both reservoirs is
qualitatively identical. All the curves in figure (\ref{Figure2}$a$)
correspond to $K=100$ and that in figure (\ref{Figure2}$b$) correspond to $%
K=0.01$. In figure (\ref{Figure2}$a$), regardless of the choice of $\tau $,
there is a strict monotonous decrease in QSLT with the increasing of
rapidity which results to a near zero value at a critical value of $\alpha
_{c}$ for all $\tau $ both for Ohmic and super-Ohmic reservoir. With further
increase in rapidity, the QSLT monotonously increases reaching a saturation
value different for each choice of $\tau $. In other words, the evolution
process constantly speeds up for $\alpha <$ $\alpha _{c}$ and then speeds
down for $\alpha >\alpha _{c}$ until it becomes constant at large value of $%
\alpha $. On the other hand, for $K=0.01$ (figure(\ref{Figure2}$b$)) there
is no decelerating effect in the evolution process. 
The QSLT for every choice of $\tau $ decreases monotonously until it reaches
a nonvanishing minimum constant value that results in uniform evolution
process.

\begin{figure}[h]
\begin{center}
\subfigure[]{
\includegraphics[scale=0.9]{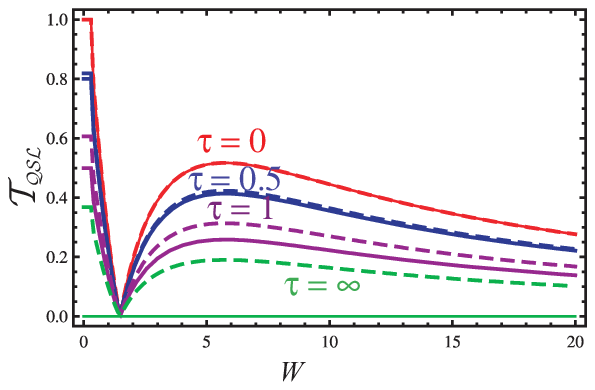}} 
\subfigure[]{
\includegraphics[scale=0.9]{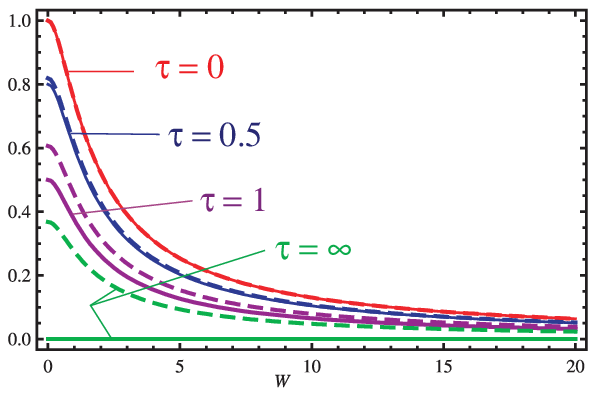}}
\end{center}
\caption{(Color Online) The $\mathcal{T}_{\mathcal{QSL}}^{\mathcal{R}}$ as a
function of the width of the wavepacket $W$ by choosing the other parameters
such that $\Delta \protect\tau =1,$ $\protect\eta =1,$ $\protect\omega %
_{c}=1,$ $\protect\alpha =\infty ,$ $\protect\theta =\protect\pi /4$ and $%
(a) $ $K=100$ $(b)$ $K=0.01.$ The red, the blue, the purple and the green
solid (dashed) curves correspond to the initial time parameter $\protect\tau %
=0,$ $0.5$ $,1$, $\infty $ for Ohmic (super-Ohmic) reservoir, respectively.}
\label{Figure3}
\end{figure}

The effect of the width $W$ of the wavepacket for the same values of the
time parameter $\tau $ as in figure (\ref{Figure2}) on the QSLT is shown in
figure (\ref{Figure3}). Again figure (\ref{Figure3}$a$) corresponds to $%
K=100 $ and figure (\ref{Figure3}$b$) to $K=0.01$. Similarly, the solid and
the dashed curves in both subfigures, respectively, represent the influence
of Ohmic and super-Ohmic reservoirs. As in figure (\ref{Figure2}$a$), here
in figure (\ref{Figure3}$a$) also exists a critical value $W_{c}$ of the
width of the wavepacket at which the QSLT, regardless of the value of $\tau $%
, reduces to a nonvanishing minimum value and then start increasing with the
increasing value of $W$. However, the comparison of the two figures
immediately reveals that in figure (\ref{Figure3}$a$) the fall in QSLT in
the region $W<W_{c}$ is sharper than in figure (\ref{Figure2}$a$).
Similarly, in the region $W>W_{c},$ the behavior of QSLT is completely
different than in figure (\ref{Figure2}$a$). Here is no saturation point,
rather, beyond $W_{c}$ there is a relatively slower increase, reaching a
maximum that happens at the half of its initial value and then falls
gradually to a nonvanishing value for each choice of $\tau $. Unlike the
effect of $W$ for large $K$, its effect for small $K$ (figure(\ref{Figure3}$%
b $)) is to relatively slow down the decrease in QSLT as compared to figure(%
\ref{Figure2}$b$).

\section{Conclusion}

We investigate the effect of relative motion on the dynamics of quantum
speed limit time for a single free spin$-1/2$ particle initially in a
superposition state and is coupled to an Ohmic-like dephasing environment.
In particular, the effects of rapidity, normalized momentum, the size of the
wavepackets and the initial time parameter in the Markovian regime both for
Ohmic and super-Ohmic reservoirs are considered. It is found that in the
presence of relative motion, the coupling with the Ohmic reservoir
constantly speeds up the evolution process without having an upper bound
there by reducing the QSLT to zero as the evolution time increases, whereas
the coupling with super-Ohmic reservoir has an upper limit on the speed at
which the evolution process proceeds uniformly. The effect of rapidity on
QSLT is not alike for the whole range of the average momentum. In the range
of small $K$, it decreases monotonically to a constant minimum. On the other
hand, in the range of large $K$ a critical value of rapidity exists at which
it reduces to a nonvanishing minimum and then increases back until it
becomes stationary. Although quantitatively different, the effect of the
width of the wavepackets on QSLT is qualitatively parallel in some respects
to that of the rapidity both in the small and the large ranges of the values
of $K$ in the intermediate range of values of the two parameters.
Nevertheless, in the limit of large values the analogy breaks, such that in
the case of $W$ it again decreases after reaching a maximum. The results of
our study may prove useful for exploring the speed of evolution of more
complex quantum systems consists of many marginal system in the frame work
of relativity that can be used in quantum information processing in the
presence of noisy environment.

\end{document}